\documentstyle[prl, twocolumn, aps, psfig]{revtex}

\def\be{\begin{equation}}
\def\ee{\end{equation}}
\def\ba{\begin{eqnarray}}
\def\ea{\end{eqnarray}}
\begin{document}
\draft
\preprint{}
\title{Solving the Initial Value Problem of two Black Holes }

\author{Pedro Marronetti and Richard A. Matzner}

\address{Center for Relativity, The University of Texas at
Austin, TX 78712-1081}

\date{\today}
\maketitle
\begin{abstract}

We solve the elliptic equations associated with the Hamiltonian and
momentum constraints, corresponding to a system composed of two black 
holes with arbitrary linear and angular momentum. These new solutions 
are based on a Kerr-Schild spacetime slicing which provides more
physically realistic solutions than the initial data based on
conformally flat metric/maximal slicing methods. The singularity/inner 
boundary problems are circumvented by a new technique that allows 
the use of an elliptic solver on a Cartesian grid where no points are 
excised, simplifying enormously the numerical problem.

\end{abstract}

\narrowtext

\section{Introduction and Method}

Any code designed to evolve a general relativistic system will
start with a solution to the initial value problem corresponding to the
astrophysical situation of interest. In the 3+1 formulation of general 
relativity, the problem is defined by the Hamiltonian and momentum 
constraints \cite{Footnote0}
\ba
R+\frac{2}{3} K^2 - A_i^{~j} A_j^{~i}  =  0 \nonumber\\
\nabla_j A_{i}^{~j} - \frac{2}{3} \nabla_i K  =  0~~.
\label{constraints}
\ea
Above, $R$ is the 3-dimensional Ricci scalar constructed from the
spatial 3-metric $g_{ij}$, and $K$ and $A_{ij}$ are the trace and 
trace-free parts of $K_{ij}$ such as
\ba
K_{ij} = A_{ij} +\frac{1}{3} g_{ij} K ~~. \nonumber
\ea
The problem of solving the Hamiltonian and momentum constraints for two
black holes has been addressed in the past by several groups
(see \cite{Cook:2000jt} and references therein). For us, it is inherently 
3-dimensional since we want to be able to specify arbitrary boost and 
spin directions for each hole. The methods in most frequent use 
\cite{Cook:1993jt,Pfeiffer:2000um} are based on an approach which chooses 
maximal spatial hypersurfaces ($K=0$) and takes the spatial 
3-metric to be conformally flat ($g_{ij} = \phi^4 \delta_{ij}$,
$\phi$ being the conformal factor). Under these conditions, the 
Hamiltonian constraint can be decoupled from the momentum constraint.
Analytical solutions for the extrinsic curvature $K_{ij}$ for holes with 
specific linear momenta can be found \cite{Kulkarni:1983}. This 
simplification leaves only one elliptic equation for $\phi$, which is 
derived from the Hamiltonian constraint. The inner boundaries (the 
throats of the black holes) are usually dealt with by imposing an isometry 
condition between two identical asymptotically flat spatial slices, 
joined by an Einstein-Rosen bridge, though other boundary conditions are 
sometimes used. Unfortunately, the numerical solution of the equation 
for $\phi$ presents a technical challenge at the inner boundaries. 
Brandt and Br\"{u}gmann \cite{Brandt:1997tf} simplified this problem 
by compactifying the internal asymptotically flat regions to obtain a 
domain without inner boundaries. However, the main disadvantage of these 
methods is not numerical, but related to the physical interpretation of 
black-hole spaces described through a conformally flat 3-metric. 
There are no space slices for which the spatial 3-metric of a single 
Kerr (non-zero spin) black hole can be written in a conformally flat way 
\cite{Garat:2000pn}, and recent work on sequences of initial data sets
for circular orbits casts some doubt on the physical realism of 
conformally flat approaches to black hole binaries \cite{Pfeiffer:2000um}.
One way to overcome this problem is to use a Kerr-Schild
slicing of spacetime \cite{Bishop:1998my}.
Matzner, Huq, and Shoemaker \cite{Matzner:1999pt} proposed a method that 
bases the initial data on a background metric and extrinsic 
curvature that is a superposition of Kerr black hole metrics written in 
ingoing Eddington-Finkelstein coordinates. Thus, for a single Kerr black 
hole one obtains the exact solution to the problem. Marronetti 
{\it et al.} \cite{Marronetti:2000rw} added to this method a variation 
on the background fields that eliminates the inner boundary 
problem, greatly simplifying the numerical treatment of the elliptic 
equations. The {\it background} fields generated in this way are also 
very good approximate solutions to the initial value problem (i.e., the 
violation of the constraints (\ref{constraints}) is small enough to fall 
below the numerical truncation error for a wide range grid spacings).

We present here the first solution to the initial value problem for 
black-hole binaries based on physically realistic background 3-metric 
and extrinsic curvature. We study the particular case of a hyperbolic 
encounter of two Kerr black holes of mass $m$, separated by a distance of 
$11.5~m$.


The method begins by specifying a conformal spatial metric which is a 
straightforward superposition of two Kerr-Schild single hole (spatial) 
metrics:
\ba
\tilde{g}_{ij} &=& \delta_{ij}
	    + 2~{}_{1}B{}_{~1}H_{~1}l_{i~1}l_{j}
	    + 2~{}_{2}B{}_{~2}H_{~2}l_{i~2}l_{j}~~,\nonumber\\
\tilde{K} &=& {}_{1}B{}_{~1}K_i^{~i}+{}_{2}B{}_{~2}K_i^{~i}~~,\nonumber\\
\tilde{A}_{ij} &=& \tilde{g}_{n(i}~~({}_{1}B{}_{~1}K_{j)}^{~n}
	+{}_{2}B{}_{~2}K_{j)}^{~n}
	- \frac{1}{3} \delta_{j)}^{~n} \tilde{K})~~,
\label{att_fields}
\ea
where the parenthesis in the subscripts denote symmetrization in $i$ and $j$.
The scalar function $H$ has a known form \cite{Matzner:1999pt}, and 
${ l_{\lambda}}$ is an ingoing null vector congruence associated with the 
solution. The fields marked with the pre-index $1$ ($2$) correspond to an 
isolated black hole with specific angular momentum $\bf{a_1}$ ($\bf{a_2}$) 
and boosted with velocity $\bf{v_1}$ ($\bf{v_2}$). The trace-free part
of the extrinsic curvature tensor $\tilde{A}_{ij}$ is also constructed from 
the superposition of the curvatures ${}_1K_i^j$ and ${}_2K_i^j$ 
associated with the black holes 1 and 2 respectively. The attenuation 
function ${}_{1}B{}$ (${}_{2}B{}$) is unity everywhere except in the 
vicinity of hole 2 (hole 1) where it rapidly vanishes, so the fields there 
are effectively those of a single black hole, thus providing an exact 
solution to the constraints for distances arbitrarily close to the 
singularities \cite{Marronetti:2000rw}.

Following the conformal decomposition presented by York and collaborators
\cite{YP}, we relate the physical metric $g_{ij}$ and the trace-free part of 
the extrinsic curvature $A_{ij}$ to the background fields through a conformal
factor:
\ba
g_{ij} &=& \phi^{4} \tilde{g}_{ij} \nonumber\\
A^{ij} &=& \phi^{-10} \tilde{A}^{ij}~~.
\label{conf_fields}
\ea
In order to find a solution to the four constraint Eqs. (\ref{constraints}), 
we add a longitudinal part to the extrinsic curvature $A^{ij}$:
\ba
A^{ij} \equiv \phi^{-10} (\tilde{A}^{ij} + \tilde{(lw)}^{ij})~~,
\ea
where $w^i$ is a vector potential to be solved for and
\ba
\tilde{(lw)}^{ij} \equiv \tilde{\nabla}^{i} w^{j} + \tilde{\nabla}^{j} w^{i} 
	- \frac{2}{3} \tilde{g}^{ij} \tilde{\nabla_{k}} w^{k}~~.
\label{lw}
\ea
Plugging Eqs. (\ref{att_fields}-\ref{lw}), into the Hamiltonian and momentum 
constraints (\ref{constraints}), we obtain four coupled elliptic equations 
for the fields $\phi$ and $w^i$ \cite{YP}:
\ba
\tilde{\nabla}^2 \phi &=&  (1/8) \big( \tilde{R}\phi 
	+ \frac{2}{3} \tilde{K}^{2}\phi^{5} -   \nonumber \\
	& & \phi^{-7} (\tilde{A}{^{ij}} + (\tilde{lw})^{ij})  
 	    (\tilde{A}_{ij} + (\tilde{lw})_{ij}) \big)  \nonumber \\ 
\tilde{\nabla}_{j}(\tilde{lw})^{ij} &=& \frac{2}{3} \tilde{g}^{ij} \phi^{6} 
	\tilde{\nabla}_{j} K - \tilde{\nabla}_{j} \tilde{A}{^{ij}} 
\label{ell_eqs}
\ea

To solve the elliptic Eqs. (\ref{ell_eqs}) we use an adaptation of an 
multigrid elliptic solver developed for the solution of the initial 
value problem of neutron-star binaries \cite{Marronetti:1998xv}, which 
uses second order stencils. To optimize the use of memory, this solver 
was designed to handle only identical black holes; only one object is 
placed on the numerical grid, the second is accounted for using reflective 
boundary conditions at the corresponding surface of the grid cube (see 
\cite{Marronetti:1998xv} for details). The four elliptic equations are 
solved consecutively in an iteration loop. The iteration starts with a 
trivial initial guess, namely 
\ba
\phi(x^i)&=&1 \nonumber\\
w^i(x^i)&=&(0,0,0) ~~~~ \forall x^i~~,
\label{i_guess}
\ea
and it relaxes the fields until the variation from one iteration step to 
the next falls below some fraction of the truncation error. This 
typically takes around 10 to 20 iteration steps. The sources for these 
elliptic Eqs. consist mostly \cite{Footnote3} of the residuals of the 
constraint equations as defined in \cite{Marronetti:2000rw}. The use of 
attenuation functions eliminates the singular behavior of these sources 
near the ring singularities, simplifying the numerical implementation of 
the elliptic solver. Instead of using excision techniques around the 
singularities \cite{Brandt2000}, we modified a standard elliptic solver 
to ``ignore'' (i.e. leave the fields with the initial values 
(\ref{i_guess})) the grid points in a small volume embedding the 
singularity. This ``inner'' region is defined monitoring the value of 
the background 3-dimensional Ricci scalar \cite{Footnote4}, and is 
chosen to be smaller than the excision volume used in current numerical 
simulations \cite{Brandt2000}, making the initial value data suitable 
for the evolutionary codes. The choice of the ``inner'' region values 
(\ref{i_guess}) is based on the fact that arbitrarily close to the ring 
singularity, the background fields recover the single-hole values 
which are an exact solution to Eqs. (\ref{constraints}). We use Robin
conditions for the outer boundaries \cite{YP}, guaranteeing that
$\phi$ and $w^i$ fall off as $constant+O(1/r)$ at infinity. 

The calculations presented in this paper were performed using 16 
processors of an Origin2000 system located at the National Computational 
Science Alliance center in Urbana-Champaign, taking close to 200 CPU hours.

\section{Results and Conclusions}

The method described above was applied to the problem of two black
holes with mass $m$ in a hyperbolic encounter configuration with 
parameters ${\bf{r_1}} = (5.75~m,0,0)$, ${\bf{v_1}} = (0,0.5~c,0)$,
and ${\bf{a_1}} =(0,0,0.5~m)$ corresponding to black hole 1 coordinate
position, velocity, and specific angular momentum. The parameters for 
black hole two are ${\bf{r_2}} = -{\bf{r_1}}$, ${\bf{v_2}} = -{\bf{v_1}}$,
and ${\bf{a_2}} = {\bf{a_1}}$. To test the convergence of the code, two 
grid resolutions were used: one with grid spacing $h=m/4$ (low) and 
another with $h=m/8$ (high). The number of grid points was 60,750 and 
536,238 respectively and the bounding box of the numerical grid 
covers a rectangle from $(-11.5~m,-5.75~m,-5.75~m)$ to 
$(11.5~m,5.75~m,5.75~m)$. We estimated the ADM mass corresponding to
the background fields and to the full numerical solution, obtaining
$2.34~m$ (note that $(1-v^2)^{-1/2} = 1.155$) and $2.18~m$ respectively. 
This was done by integrating over the boundary surface of the computational 
grid. The difference is accounted for by the binding energy. We also 
obtained a background total angular momentum estimation of $7.99~m^2$, 
consistent with analytical estimates.

\begin{figure}[htb]
\begin{center}
\hskip 0.5 cm
\mbox{\psfig{figure=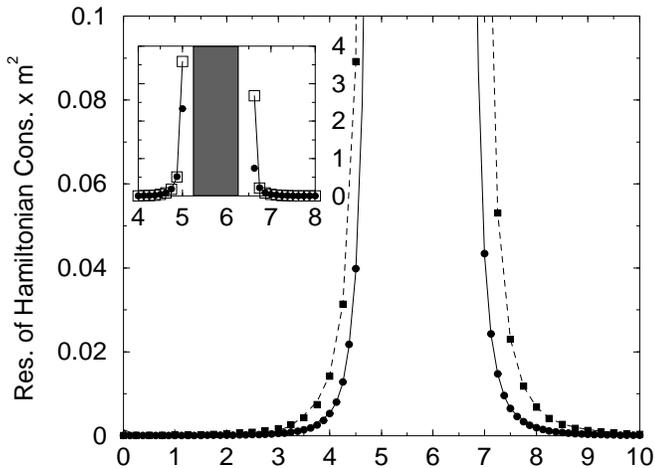,width=8 cm,height=3.0 cm}}
\end{center}
\vskip 3.8 cm
\caption{Residuals of the Hamiltonian constraint near the hole 
located at $(5.75~m,0,0)$. 
The full (dashed) line corresponds to the two-hole calculation done 
with grid spacing $h=m/8$ ($h=m/4$). The circles (squares) correspond 
to the computation of the Hamiltonian constraint of a single black hole 
on a grid with spacing $h=m/8$ ($h=m/4$).}
\label{HC}
\end{figure}

\begin{figure}[htb]
\begin{center}
\hskip 0.5 cm
\mbox{\psfig{figure=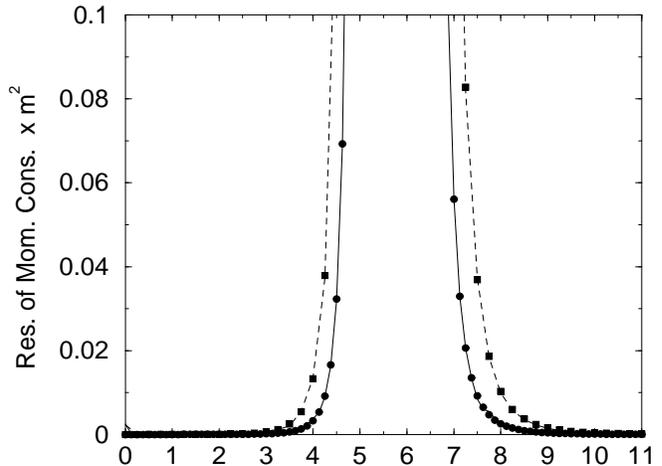,width=8 cm,height=3.0 cm}}
\end{center}
\vskip 3.8 cm
\caption{Residuals of the Momentum constraint (x comp.) near 
the hole located at $(5.75~m,0,0)$. The notation is the same as in 
Figure \ref{HC}.}
\label{MCx}
\end{figure}

When differenced under our numerical scheme, even analytically exact
single-hole solutions show inevitable nonzero residual errors in the
Hamiltonian and momentum constraints, which increase near the hole, due
to the larger gradients there. In Figures \ref{HC} and \ref{MCx}
these are plotted as points  (filled circles for resolution $m/8$, 
filled squares for resolution $m/4$). The  decrease of this truncation 
error with increasing resolution shows the convergent behavior of the 
differencing scheme.  Also in Figures 1-2 we plot as curves the 
computed constraint residuals of the rightmost hole in our two-hole 
solution. Except for the innermost points in the solution, the residuals 
are effectively identical to those of the analytical single-hole 
solution at the same resolution. The inset in Figure \ref{HC} shows this 
discrepancy at the innermost points for the high resolution curve
(empty squares mark the grid points and the singularity
is represented by a gray vertical band). We thus confirm our two-hole 
solution at the level of truncation error except for these innermost 
points, which will anyway be excised in the evolution scheme
\cite{Footnote5}.

\begin{figure}[htb]
\begin{center}
\hskip 2.0 cm
\mbox{\psfig{figure=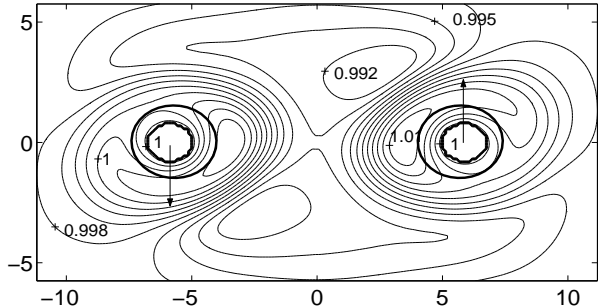,width=8 cm,height=4.3 cm}}
\end{center}
\vskip 0.5 cm
\caption{Contour plot of the conformal factor $\phi$. 
The inner thick lines limit the ``inner''
region where the initial value $\phi=1$ is preserved. The outer thick
lines show the apparent horizons and the arrows the direction of the 
boosts. }
\label{phi}
\end{figure}

Figure \ref{phi} is a contour plot of the conformal factor $\phi$. 
The inner thick lines limit the ``inner'' region where the initial value 
$\phi=1$ is preserved. The apparent horizons of the holes were calculated 
using a code developed by Huq {\it et al.} \cite{Huq:2000qx} and are 
represented by the outer thick lines. The resulting horizon areas for the 
background system and the full solution were $2 \times 39.9~m^2$ and 
$2 \times 38.4~m^2$ respectively, which are consistent with the 
value of $45.7~m^2$ obtained for a single-hole configuration .

Table \ref{TableI} shows the $l_1$ (average of the absolute value) and 
$l_\infty$ (maximum value) norms for the calculated fields. 
The final fields $\phi$ and $w^i$ depart very little from the initial guess 
(\ref{i_guess}) all across the numerical grid, indicating the great 
degree of accuracy of the background fields as approximate solutions to 
the constraints \cite{Marronetti:2000rw}. 

The attenuation method, while providing a clear numerical advantage, also 
raises the question of physical interpretation of these results. One would
expect the fields around each hole to be tidally distorted by the presence
of the second hole, and the attenuation functions partially occlude this 
effect. However, we notice that most of this effect is confined to the 
regions very near the ring singularities: i.e. regions that will be excised 
from the computational domain by the time evolution code. At the same time,
we see from figure \ref{phi} that the tidal interaction is still 
(at least partially) present at the location of the apparent horizons.
The influence of the attenuation functions can be interpreted as 
an addition of gravitational radiation that conveniently (from the 
numerical point of view) eliminates the singular behavior from the 
elliptic equations near the singularities. 

This solution to the initial value problem presents a different approach 
than the conformally flat/maximal slicing methods, and allows us to 
specify more directly the physical content of the data. It also entails 
a computational simplification on the treatment of the singularities on 
the working grid. However, the physical aspects of these data sets can 
be further refined, for instance by providing background fields based on 
Post-Newtonian expansions or with a more physical 
prescription for the background fields for circular orbits \cite{MG9}.


This work was supported by NSF grant PHY9800722 and PHY9800725 to 
the University of Texas at Austin. Most of the elliptic solver was 
developed as part of the thesis work of one of the authors (PM) at the 
University of Notre Dame. PM wants to express his gratitude to 
Grant J. Mathews and James R. Wilson for their guidance and 
encouragement. Special thanks to L. Lehner for his help with the 
apparent horizons calculations and to M. Huq, P. Laguna, D. Nielsen, 
D. Shoemaker, and G. B. Cook for very helpful discussions.

\begin{table}
\caption{Maximum ($l_\infty$) and average of the
absolute values ($l_1$) of the conformal factor $\phi$ and vector
potential $w^i$ on the grid volume.}
\begin{tabular}{lcccc}
\  & $\phi$ & $w^x$ & $w^y$ & $w^z$ \\
\tableline
&&&&\\
$l_\infty$    &   $1.009e+0$ & $0.462e-2$ & $0.168e-1$ & $0.421e-2$\\
$l_1$         &   $0.997e+0$ & $0.745e-3$ & $0.146e-2$ & $0.452e-3$\\
\end{tabular}
\label{TableI}
\end{table}


\begin{thebibliography}{99}
\bibitem{Footnote0}  
Covariant differentiation with respect to $g_{ij}$ ($\tilde{g}_{ij}$)
is denoted by ${\nabla}_i$  ($\tilde{\nabla}_i$). 
Spacetime indices will be denoted by Greek letters and spatial 
indices by Latin letters.
\bibitem{Footnote3} 
In the numerical implementation of the solver, the terms on the rhs. of 
Eqs. (\ref{ell_eqs}) that include the background affine connection 
coefficients are passed to the sources. These terms still have the
``singularity'' embedded in them, thus part of the singular behavior 
remains. Fortunately, the numerical experiments showed that their 
influence in the convergence of the algorithm is very small and confined 
to the points closest to the inner region limit.
\bibitem{Footnote4} 
If the Ricci scalar exceeds a given threshold value $R_t$, then
the grid point is inside the ``inner'' region. For this particular 
hyperbolic encounter, we used $R_t=2.0/m^2$.
\end{thebibliography}
\end{document}